\documentstyle{article}

\def\e{\epsilon}
\font\tit=cmr10 scaled\magstep4
\begin{document}
\input{epsf}
\centerline{\tit Action functionals for strings in four dimensions}
\vskip 1.5truecm
\centerline{\bf Mikhail Alexandrov}
\medskip
\centerline{\sl University of California}
\centerline{\sl Department of Mathematics}
\centerline{\sl Davis, CA 95616}
\medskip
\centerline{\sl E-mail: alexandr@ucdmath.ucdavis.edu}
\vskip 2truecm
\centerline{\bf Abstract}
\medskip

{\sl All possible action functionals on the
space of surfaces in ${\bf R}^4$ that depend only on first and second
derivatives of the functions, entering the equation of the surface,  and
satisfy the condition of invariance with respect to rigid motions are
described.}

\vskip 2truecm

It was conjectured that quantum chromodynamics is equivalent  to
some kind of string theory.
The bosonic part of the action functional of such a string theory
should be considered as a functional defined on the space of  2-dimensional
surfaces in 4-dimensional space.
The simplest possible action (Nambu-Goto action = area of the surface in
the induced metric) can not describe the QCD string; A. Polyakov [1]
suggested to include in the string action a term expressed in terms of
extrinsic curvature of the surface in the induced metric.

In the present paper we describe all possible action functionals on the
space of surfaces in ${\bf R}^4$ that depend only on first and second
derivatives of the functions, entering the equation of the surface,  and
satisfy the natural condition of invariance with respect to rigid motions.
Similar results can be obtained for action functionals defined on the space
of surfaces in Minkowski space.

\medskip

It is assumed that the action functional can be represented in the form:

$$S = \int\limits_{\Gamma} A \left(X(U),
{\partial X^\alpha \over \partial U^j},
{\partial^2 X^\alpha \over {\partial U^j \partial U^k}} \right) dU. \eqno(1)$$
Here we denote by $X^\alpha$ the coordinates in ${\bf R}^4$ and by $U^j$ the
parameters of the 2-dimensional surface $\Gamma$.
Of course, the integral (1) should be independent
on the choice of parameterization  $X=X(U)$ of $\Gamma$.
In this case the function $A$ is called a {\sl density}.
We supposed that the function $A$ depends on the first and second
derivatives of $X(U)$ only; in this case one says that $A$ is a density of
{\sl rank 2}.
We assume the invariance of $S$ with respect to  rigid motions of ${\bf R}^4$.
In particular, $S$ should be invariant with respect to
shifts (parallel transports), therefore  $A$
will not depend on $X(U)$ itself but only on its derivatives.

There are two possible ways to define the measure of
integration $dU$ in the integral (1): as a positive measure,
or as a signed one (it changes sign under a orientation reversing
reparametrization).
In the second case we will deal with oriented surfaces only.
The definition of density above requires $A$ to be "covariant" under the
action of the reparametrization group $L$.
And for the different definitions of $dU$ the meaning of the word "covariant"
is not the same: in the case of a reparametrization $U=K(U')$ ($K \in L$)
the new density $A'$ can be obtained from the old one $A$ by the
multiplication either by
the Jacobian of the transformation $K$ (density of the {\it second kind}), or
by the absolute value of this Jacobian (density of the {\it first kind}).
Since the results in these two cases are very similar, we will consider
only densities of the first kind.

\medskip

We will prove that all densities satisfying the requirements above can be
expressed in terms of the following four basic invariants:
$$Q_1 = {1 \over \Omega} \e^{km} \e^{ln} \delta_{\alpha \beta}
D_k \partial_l X^\alpha  D_m \partial_n X^\beta, \eqno(2)$$
$$Q_2 = {1 \over \Omega^2} \e^{ik} \e^{jn} \e^{pl} \e^{qm} \delta_{\alpha
\beta}
\delta_{\gamma \tau} \delta_{\mu \nu}
\partial_i X^\gamma \partial_j X^\tau
\partial_p X^\mu \partial_q X^\nu
D_k \partial_l X^\alpha D_m \partial_n X^\beta, \eqno(3)$$
$$Q_3 = {1 \over \Omega^2} \e^{pq} \e^{ik} \e^{jl} \e^{mn}
\e_{\alpha \beta \gamma \delta} \delta_{\mu \nu}
\partial_i X^\mu \partial_j X^\nu
\partial_m X^{\alpha} \partial_n X^{\beta}
D_p \partial_l X^{\gamma} D_k \partial_q X^{\delta}. \eqno(4)$$
$$Q_4 = {1 \over \Omega^2} \e^{im} \e^{kp} \e^{lq} \e^{jn}
\delta_{\alpha \beta} \delta_{\mu \nu}
D_i \partial_j X^\alpha D_k \partial_l X^\beta
D_p \partial_q X^\mu  D_m \partial_n X^\nu. \eqno(5)$$
Here $D_i \partial_j X^\alpha$ are covariant derivatives of the corresponding
1-forms (in parameter space) $\partial_j X^\alpha$ (for fixed $\alpha$)
with respect to the metric
$s_{ij} = \delta_{\alpha \beta} \partial_i X^\alpha \partial_j X^\beta$
induced by the embedding of the surface $\Gamma$ in ${\bf R}^4$, and
$$\Omega = {1 \over 2} \e^{ik} \e^{jl} \delta_{\alpha \beta} \delta_{\mu \nu}
\partial_i X^\alpha \partial_j X^\beta \partial_k X^\mu \partial_l X^\nu
\eqno(6)$$
is the determinant of $s$.

To construct a density one should take a real function in four variables
$f(p, q, r, t)$ that is invariant under the transformation $r \rightarrow -r$
and is regular in the domain specified by the conditions:
$p \ge 0$, $q \ge 0$, $t \ge 0$, $p^2 > 4 r^2$, $pqt \ge q^2 r^2 + t^2$;
then the corresponding density will have the form:
$$A = F(Q_1,Q_2,Q_3,Q_4)\  \sqrt{\Omega}, \eqno(7)$$
where
$$F(Q_1,Q_2,Q_3,Q_4) = f\left({Q_2 - Q_1 \over 4}, {Q_2 + Q_1 \over 4},
{Q_3 \over 8}, {Q_1^2 + Q_2^2 \over 32} - {Q_3^2 \over 128} -
{Q_4 \over 8}\right). \eqno(8)$$

We will prove that {\sl every reparametrization invariant action functional of
the form (1) can be obtained by means of the above construction}.

If $F$ is identically equal to $1$, we have the usual {\sl Nambu-Goto} density,
having the meaning of the surface area; $F=Q_1$ corresponds to the
{\sl Einstein action functional} (here density is equal to the scalar intrinsic
curvature of the surface); if $F=Q_2$ we obtain the extrinsic curvature density
[1].

\medskip

The proof of these results is based on the ideas of [2, 3].
It is pointed out there, that the density $A$ can be considered as a function
on the space ${\cal Q}$ of all quadratic surfaces, i.e. surfaces of the form:
$$x = B_i u^i + G_{ij} u^i u^j.\eqno(9)$$
Really, the density $A$ in (1) depends only on the first two derivatives of $X$
and for each point $X_0$ of $\Gamma$ (corresponded to the value $U_0$ of the
parameter) we can consider these derivatives as coefficients
of a quadratic surface $\Sigma_{X_0}$, approximating $\Gamma$ at this point:
$$B_i = {\partial X \over \partial U^i}\bigg|_{X_0} =
{\partial x \over \partial u^i}\bigg|_0, \ \ \
G_{ij} = {1 \over 2} {\partial^2 X \over \partial U^i \partial U^j}\bigg|_{X_0}
 = {1 \over 2} {\partial^2 x \over \partial u^i \partial u^j}\bigg|_0.$$
Here $u = U-U_0, x = X - X_0$.
We see that $G$ is a symmetric $2 \times 2$ matrix with values in ${\bf R}^4$
and $B$ is a $1 \times 2$ matrix also with values in ${\bf R}^4$.
Sometimes we will use the more explicit form of this formula:
$$x = a u + b v + f u^2 + 2g u v + h v^2 , \eqno(9a)$$
where $u=u^1, v=u^2$, $a=B_1, b=B_2, f=G_{11}, g=G_{12}=G_{21}, h=G_{22}$
and all the coefficients are vectors in ${\bf R}^4$.

To describe densities as functions on ${\cal Q}$ we consider the group
$\Lambda$ acting on ${\cal Q}$ and generated by linear and quadratic
reparametrizations.

The {\sl linear reparametrization}
$$u^i \rightarrow K^i_j u^j \eqno(10)$$
transforms the surface (2) into the surface
$$x = B'_i u^i + G'_{ij} u^i u^j,$$
where
$$B'_i = K^j_i B_j, \ \ \ G'_{ij} = K^l_i K^m_j G_{lm}.$$
The {\sl quadratic reparametrization}
$$u^i \rightarrow u^i + T^i_{jk} u^j u^k \eqno(11)$$
transforms (2) into the quadratic surface
$$x = B_i u^i + G'_{ij} u^i u^j,$$
where
$$G'_{ij} = G_{ij} + B_k T^k_{ij}.$$
It is easy to show that
the density of  rank 2 can be considered as a function $A$ defined on the
space ${\cal Q}$ that is invariant under quadratic reparametrizations and  its
value after linear reparametrization (10) can be obtained from the initial one
by means of multiplication on $|{\sl det}K|$.

\medskip

It is not difficult to show, that
any surface $\Sigma \in {\cal Q}$ can be reduced to some {\sl standard form}
$\Sigma^0$:
$$x = \pmatrix{{1}\cr {0}\cr {0}\cr {0}\cr}u +
\pmatrix{{0}\cr {1}\cr {0}\cr {0}\cr}v +
\pmatrix{{0}\cr {0}\cr {\alpha + \beta}\cr {\gamma}\cr}u^2 +
2\pmatrix{{0}\cr {0}\cr {0}\cr {\mu}\cr}uv +
\pmatrix{{0}\cr {0}\cr {\beta - \alpha}\cr {\gamma}\cr}v^2$$
by means of transformations from the group $G = SO(4) \times \Lambda$.
Here $\alpha, \beta, \gamma,$ and $\mu$ are some arbitrary real numbers.
This form is {\sl almost} unique: two standard forms that differs only by the
sign of $\alpha$ correspond to the same surface $\Sigma$.
So, we will consider $\alpha^2$ as a parameter of $\Sigma^{(0)}$ instead of
$\alpha$ itself.

We should notice, that the properties of density described above allow us
to reconstruct the value of a density on a surface $\Sigma  \in {\cal Q}$ from
its value on the corresponding surface $\Sigma^0$ of the standard form.
It follows also from these properties, that the ratio of two densities is
invariant with respect to the action of $G$.
It will be convenient for us to represent densities in the form
$F \sqrt{\Omega}$,
where $F$ is an invariant with respect to $G$, and $\sqrt{\Omega}$ is the
simplest density of rank 2 ($\Omega$ is defined by (6)).

The parameters $\alpha^2, \beta, \gamma,$ and $\mu$ of the surface in the
standard form  are {\sl invariants} of the surface $\Sigma$ under the action of
the
group $G$.
This means that, being expressed through the coefficients
$B_i$ and $G_{ij}$  they are not be changed under transformations
from the group $G$.
Any other invariant can be expressed in terms of these invariants
(since it can be expressed on the standard surface).
We see that the total number of independent invariants
(= number of independent parameters of standard surface) is equal to 4.
We will use the word {\sl basis} to denote a minimal set of invariants such
that any other invariant can be represented as a function of the basic ones.
It is clear that any function of invariants is an invariant
itself, and two invariants  are identical if they coincide on the standard
surface.

It is convenient for further use to modify the basis of invariants on the
standard surface:
$$p = \alpha^2 + \mu^2, \ \ \ q = \beta^2 + \gamma^2,$$
$$r^2 = \alpha^2 \mu^2, \ \ \ \ t = \alpha^2 \beta^2 + \gamma^2 \mu^2,$$
this basis is equivalent to the original one when $p \ge 0$, $q \ge 0$,
$t \ge 0$, $p^2 > 4 r^2$ and $pqt \ge q^2 r^2 + t^2$.

We will consider expressions having the form of inner products
of invariant tensors
$\delta_{\alpha \beta}, \epsilon_{\alpha \beta \gamma \mu}, \epsilon^{ij}$
with covariant derivatives of $x$ with respect to parameters $u=u^1, v=u^2$
(or of $X$ with respect to $U^1$ and $U^2$):

$$A_i^\alpha = \partial_i x^\alpha \big|_0 = \partial_i X^\alpha \big|_{X_0},
\ \ \ F_{ij}^\alpha = {1 \over 2} D_i \partial_j x^\alpha \big|_0 =
{1 \over 2} D_i \partial_j X^\alpha \big|_{X_0},$$
$$i,j = 1,2; \ \ \alpha = 1,2,3,4$$
(inner product is defined by means of contraction of indices,
Greek indices correspond to coordinates in
${\bf R}^4$, and Roman indices denote coordinates in the 2-dimensional
parameter space).
$F_{ij}^\alpha$ are covariant derivatives of ${1 \over 2} \partial_j x^\alpha$
which can be considered as 1-forms in parameter space for any fixed $\alpha$.
It is easy to check that $A_i^\alpha$ and $F_{ij}^\alpha$ are covariant with
respect to linear reparametrizations (10) and invariant with respect to
quadratic reparametrizations (11); both these quantities obey vector
transformation rule under the action of $SO(4)$.
We suppose all indices in our inner products to be contracted.
It is clear that such expressions transform by some one-dimensional
representation of the group $G$.
The ratio of two expressions obeying the same transformation  rules is an
invariant  with respect to the action of $G$.
We will form a basis of invariants having this form.
To check their independence we will compare their values on the standard
surface with the basic invariants obtained above.
(Note, that by definition of covariant derivative the difference between
$F_{ij}$ and the usual
derivative $G_{ij}$ from formula (9) is a linear combination of the first order
derivatives $B_i = A_i$ ($a$ and $b$ in the form (9a)) with Christoffel
symbols as coefficients.
On the standard surface Christoffel symbols at $x=0$ are equal to zero;
$\Omega=1$.
So, calculations on the standard surface are much simpler then in
generic case.)

Since our aim at this moment is not to list all possible invariants but to
find a convenient basis, we can impose some restrictions on the form of
expressions sought.
Let us consider the following tensors (with respect to the action of the group
$\Lambda$):
$$s_{ij} = <A_i, A_j>, \ \ \ H_{ij,kl}=<F_{ij}, F_{kl}>, \ \ \
T_{ij;kl,mn}=
\epsilon_{\alpha \beta \gamma \delta} A_i^\alpha A_j^\beta F_{kl}^\gamma
F_{mn}^\delta,$$
where $< \ , \ >$ denote the usual inner product in ${\bf R}^4$.
Since any inner product (with respect to all indices) of some
number of $\epsilon^{ij}$ with some of these tensors transforms by some
1-dimensional representation of $\Lambda$ and is invariant under rotations
in ${\bf R}^4$, we can divide this inner product by an appropriate power of
$\Omega = {1 \over 2} \e^{ik} \e^{jl} s_{ij} s_{kl}$ to get an invariant
with respect to $G = SO(4) \times \Lambda$.
The corresponding power of $\Omega$ is defined by the following condition:
$$ 2 \#\Omega = \#s + 2 \#H + 3 \#T. \eqno(12)$$
Here $\#s$, $\#H$ and $\#T$ denote the number of appearance of the
corresponding symbol in the expression, $\#\Omega$ is  the power of
$\Omega$ in the denominator.
It follows from these formulas, that the number of tensors
$\epsilon^{ij}$ in the expression should be equal to $2\#\Omega$.

Symmetry properties of $s, H, Y$ and $T$ follow immediately from their
definitions.

We can assign the number $n = \#H + \#T$ to each invariant of the form
above.
It plays role of a quantitieve measure of simplicity of the invariant and
will be called the {\sl order} of this invariant.

We would like to stress again that ,being invariants, the expressions sought
have the same values on every surface corresponded to the same standard form.
Then, we can calculate these values on the standard surface, where it is
easier.

\medskip

The formulas under consideration contain many indices that makes them hardly
comprehensive.
To reduce this inconvenience let us introduce the following diagram technique.
We assign the following  (vertex) diagrams:

\vspace{0.25in}
\begin{figure}[ht]
\centerline{\epsfbox{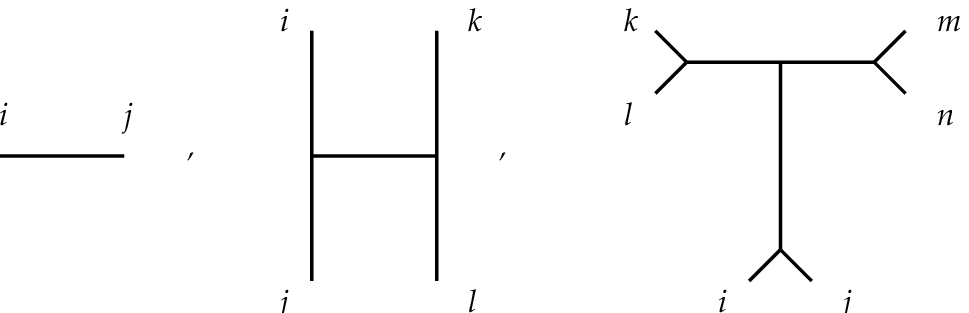}}
\end{figure}
\vspace{0.25in}

\noindent to $s_{ij}$, $H_{ij,kl}$, and $T_{ij;kl,mn}$ correspondingly.
$\epsilon^{ij}$ will be denoted by the dashed line:

\vspace{0.25in}
\begin{figure}[ht]
\centerline{\epsfbox{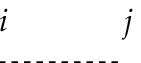}}
\end{figure}
\vspace{0.25in}

And, as in many well-known diagram techniques, contraction of indices  will be
indicated by merging of corresponding edges at the point corresponding to
the index of that contraction (we will not write contracted indices on the
graph).

For example, the diagram for $2\Omega =  \e^{ik} \e^{jl} s_{ij} s_{kl}$ will be
the following:

\vspace{0.25in}
\begin{figure}[ht]
\hspace{1in}
\centerline{\epsfbox{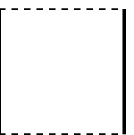}}
\end{figure}
\vspace{0.25in}

We will consider connected diagrams only.
Really, if an invariant corresponds to a disconnected diagram, then it can be
represented as a product of the invariants corresponding to its components,
that are simplier then the original one.

\medskip

Using formula (12), symmetry properties of vertices and the following equality:

\vspace{0.25in}
\begin{figure}[ht]
\centerline{\epsfbox{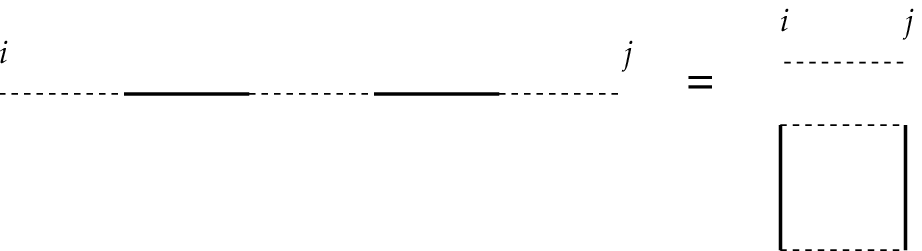}}
\end{figure}
\vspace{0.25in}

\noindent one can easily see that the number of independent invariants of
the first order can not be greater then 3.
Let us take the following three simplest diagrams of the first order:

\vspace{0.25in}
\begin{figure}[ht]
\centerline{\epsfbox{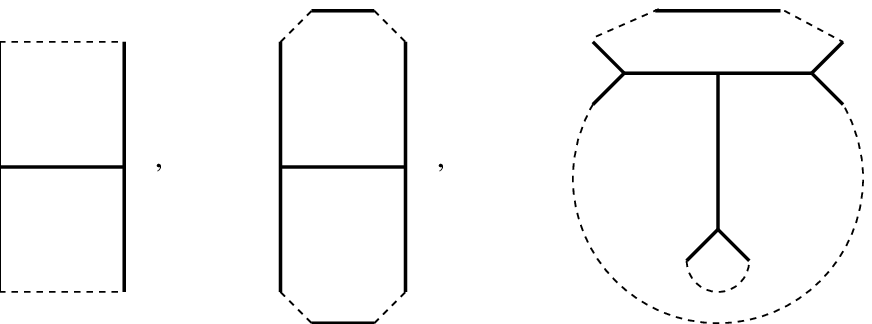}}
\end{figure}
\vspace{0.25in}

\noindent and denote the corresponding invariants by  $Q_1$, $Q_2$ and $Q_3$.
The explicit expressions for them are:
$$Q_1 = {1 \over \Omega} \e^{km} \e^{ln} <F_{kl}, F_{mn}>,$$
$$Q_2 = {1 \over \Omega^2} \e^{ik} \e^{jn} \e^{pl} \e^{qm} s_{ij} s_{pq}
<F_{kl}, F_{mn}>,$$
$$Q_3 = {1 \over \Omega^2} \e^{pq} \e^{ik} \e^{jl} \e^{mn}
\e_{\alpha \beta \gamma \delta} s_{ij} A_m^{\alpha} A_n^{\beta}
F_{pl}^{\gamma} F_{kq}^{\delta}.$$
To show that they are independent one can calculate their
values on the standard surface:
$$Q_1^s = 2[ -\alpha^2 + \beta^2 + \gamma^2 - \mu^2] = 2[q - p],$$
$$Q_2^s = 2[ \alpha^2 + \beta^2 + \gamma^2 + \mu^2] = 2[q + p],$$
$$Q_3^s = 8\alpha \mu = 8r.$$

To complete the basis we need one more invariant independent from them.
Actually, any invariant with that property can be basic.
To get the simplest one  we will take the simplest diagram of the
second order:

\vspace{0.25in}
\begin{figure}[ht]
\centerline{\epsfbox{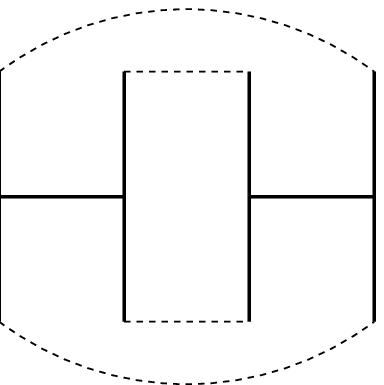}}
\end{figure}
\vspace{0.25in}

\noindent The fact that the corresponding invariant
$$Q_4 = {1 \over \Omega^2} \e^{im} \e^{kp} \e^{lq} \e^{jn}
<F_{ij},F_{kl}> <F_{pq},F_{mn}>$$
is independent from $Q_1, Q_2$ and $Q_3$ follows from its form the standard
surface:
$$Q_4^s = 4[\alpha^4 + \beta^4 + \gamma^4 + \mu^4] +
8[\beta^2 \gamma^2 - \alpha^2 \beta^2 - \gamma^2 \mu^2] =
2[q^2 + p^2 - 2r^2 - 4t].$$

\medskip

So, $Q_i, \ i=1,2,3,4$ form a basis in the space of invariants, i.e. any other
invariant can be represented as a function of $Q_i$.
For example, the invariant corresponding to the diagram

\vspace{0.25in}
\begin{figure}[ht]
\centerline{\epsfbox{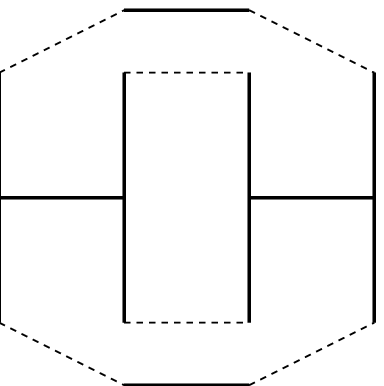}}
\end{figure}
\vspace{0.25in}

\noindent can be represented as ${1 \over 2} Q_1 Q_2 + {1 \over 16} Q_3^2$.
(It is sufficient to check this only on the standard surface.)

\medskip

The above construction proves the description of reparametrization invariant
functionals on the space of surfaces in ${\bf R}^4$ given in the begining of
the paper.

\vskip 1.5truecm

\centerline{\bf Acknowledgements}
\medskip

I am indebted to Albert Schwarz, who suggested the problem, and to Dmitri
Fuchs for useful discussions.

\vskip 2truecm
\centerline{\bf References}
\begin{enumerate}

\item A.M. Polyakov: Fine Structure of strings, Nucl.Phys.,
{\bf B268}, 406 (1986)

\item A.V. Gaiduk, V.N. Romanov, and A.S. Schwarz: Supergravity and field space
democracy, Commun.Math.Phys., {\bf 79}, 507 (1981)

\item A.V. Gaiduk, O.M. Khudaverdian, and A.S. Schwarz: Integration over
surfaces in a superspace, Theor.Math.Phys., {\bf 52}, 862 (1983)

\end{enumerate}

\end{document}